\documentclass[aps,prd,twocolumn]{revtex4} 
\usepackage{amsmath,amsfonts,amssymb}
\begin{document}
\title{\bf Massive Klein--Gordon equation from a BEC-based analogue spacetime}
\author{Matt Visser}
\email{matt.visser@mcs.vuw.ac.nz}
\author{Silke Weinfurtner}
\email{silke.weinfurtner@mcs.vuw.ac.nz}
\affiliation{School of Mathematics, Statistics, and Computer Science,
Victoria University of Wellington,
PO Box 600, Wellington, New Zealand.}

\begin{abstract}
  
We extend the ``analogue spacetime'' programme by investigating a condensed-matter system that is in principle capable of simulating the massive Klein--Gordon equation in curved spacetime. Since many elementary particles have mass, this is an essential step in building realistic analogue models, and a first step towards simulating quantum gravity phenomenology. 
Specifically, we consider the class of two-component BECs subject to laser-induced transitions between the components. This system exhibits a
  complicated spectrum of normal mode excitations, which can be viewed as two
  interacting phonon modes that exhibit the phenomenon of  ``refringence''. 
  We study the conditions required to make
  these two phonon modes decouple. Once decoupled, the two distinct phonons generically couple to distinct effective spacetimes, representing a bi-metric model, with one of the modes acquiring a mass. In the eikonal limit the massive mode exhibits the dispersion relation of
  a massive relativistic particle $\omega = \sqrt{\omega_0^2 + c^2 k^2}$, plus curved-space modifications.  Furthermore, it is possible to tune the system so that both modes can be to arranged travel at the same speed, in which case the two phonon excitations couple to the same effective metric. From the analogue spacetime perspective this situation corresponds to the Einstein equivalence principle being satisfied.   


\centerline{gr-qc/0506029}

\medskip

\centerline{6 June 2005; \LaTeX-ed \today }

\end{abstract}
\maketitle

\clearpage
\newcommand{\norm}[1]{\left\Vert#1\right\Vert}
\newcommand{\betrag}[1]{\left\vert#1\right\vert}
\newcommand{\bra}[1]{\langle #1\vert}
\newcommand{\ket}[1]{\arrowvert #1 \rangle}
\newcommand{\braket}[1]{\langle #1\rangle}
\newcommand{\kin}{-\frac{\hbar^2}{2m} \, \nabla^2 \,}
\newcommand{\oppdag}[2]{\hat{#1}^\dagger(t,\vec{#2})}
\newcommand{\opp}[2]{\hat{#1}(t,\vec{#2})}
\newcommand{\bog}{\hat{\Psi}(t,\vec{x})=\Phi(t,\vec{x}) \, + \, \varepsilon \hat{\psi}(t,\vec{x})\, + \, \ldots}
\newcommand{\zeit}{i \,\hbar \,  \frac{\partial}{\partial t} \,}
\newcommand{\mad}{\sqrt{\rho(t,\vec{x})}e^{i\theta(t,\vec{x})}e^{-\frac{i\mu t}{\hbar}}}

\section{Introduction}
Analogue models for gravitation (which should more accurately be referred to as analogue models for curved space-time) can be used to simulate classical and
quantum field theory in curved space-time~\cite{Unruh, Novello:2003, Volovik:2000, Garay, BLV-BEC, Fedichev, BLV-probe, silke2, Uwe:new, Ergo, normal, normal2, birefringence, review, book, droplet, review2, vortex}.  The first analogue model
for black holes, and for simulating Hawking evaporation, was suggested
by Bill Unruh \cite{Unruh} in 1981.  He demonstrated that a sound wave
propagating though a converging fluid flow exhibits the same
kinematics as does light in the presence of a curved space-time
background.  Since then several other media 
--- \emph{e.\,\!g.} flowing dielectrics~\cite{Novello:2003} and quantum liquids~\cite{Volovik:2000} --- 
have been analyzed, and
the field has developed tremendously~\cite{review,book,droplet}. The first approach specifically
using Bose--Einstein condensates as an analogue model was made
some nineteen years after Unruh's original paper~\cite{Garay}.  Since then various configurations
of BECs have been studied to simulate different scenarios for
gravity~\cite{BLV-BEC, Fedichev, BLV-probe, silke2, Uwe:new}.
Until now it has only been possible to simulate photons, (generally
speaking, massless relativistic particles), propagating through a
curved space-time~\cite{Ergo, normal, normal2, birefringence, review, book, droplet, review2, vortex}. In the present article a two-species BEC is used to extend the class of equations
that can be simulated to the full curved-space massive Klein--Gordon
equation in (3+1) dimensions.  (In the language of the BEC community this corresponds to a specific 
and technologically interesting 
way of giving a mass to the phonon.) From the viewpoint of the general relativity community, this article provides a route for analogue simulations of curved-space quantum field theory that are more general and realistic than those considered to date.  (A preliminary sketch of these of these ideas appeared in reference~\cite{Letter}. An alternative route to the Klein--Gordon equation via textures in 3He is described in~\cite{Textures}, and the use of ``wave guides'' to restrict transverse oscillations and provide an effective 1+1 dimensional mass is described in~\cite{Uwe:new}.)

\section{Two-component BECs}
The class of system we will use in our theoretical analysis is a two-component BEC. More specifically we consider an ultra-cold
two-component BEC atomic gas such as, for example, a two-component condensate of
${}^{87}$Rb atoms in different hyperfine levels, which we label
$\ket{A}$ and $\ket{B}$.  (Experiments using two different spin
states, $\ket{F=1,m=-1}$ and $\ket{F=2,m=2}$, were first performed at
\textit{JILA} in 1999~\cite{Myatt:1997}.)  
At very low temperatures nearly all atoms occupy the ground state.
For the following calculation the non-condensed atoms are neglected.
(In \cite{Fedichev:1998, Burnett:1998a, Burnett:1998b, Burnett:2000, Burnett:2004}
finite temperature effects are taken into account.)
The quantized field describing
the microscopic system can be replaced by a classical mean-field, a
macroscopic wave-function.  In this so-called mean-field approximation
the number of non-condensed atoms is small. Interactions between the
condensed and non-condensed atoms are neglected in the mathematical
description, but two-particle collisions between condensed atoms are
included.  In the case of a two-component system, interactions within
each species ($U_{AA}$, $U_{BB}$) and between the different species
($U_{AB}=U_{BA}$) take place.  In addition the two condensates are
coupled by a laser-beam, which drives transitions between the two
hyperfine states with a constant rate $\lambda$.  Without the laser
coupling $\lambda$, no mass term is generated, which is consistent
with the analysis in reference~\cite{Uwe:new}.
(In that article the advantages of using a two-component BEC to simulate 
cosmological inflation are
presented.) 
Because, for current purposes, the two species are different hyperfine levels of the same atom, the masses of the individual atoms are approximately equal (to about one part in $10^{16}$). In the current article we therefore set
\begin{equation}
m_A = m_B = m,
\end{equation}
 though we note that strictly speaking $m_A\neq m_B$. 
 (Indeed in more abstract situations, potentially of relevance in quantum gravity phenomenology, it is worthwhile and instructive to keep this extra mass dependence explicit.)

The resulting coupled two-component time-dependent
Gross--Pitaevskii equations are:
\begin{eqnarray}
 \label{2GPE}
 i \, \hbar \, \partial_{t} \Psi_{A} &=& \Bigg[ -\frac{\hbar^2}{2\,m} \nabla^2 
 + V_{A} - \mu_{A} 
 \\
 &&
 + U_{AA} \, \betrag{\Psi_{A}}^2  + U_{AB} \betrag{\Psi_{B}}^2 \Bigg]  \Psi_{A} 
 + \lambda \, \Psi_{B} \, ,
 \nonumber
\\
 i \, \hbar \, \partial_{t} \Psi_{B} &=& \Bigg[ -\frac{\hbar^2}{2\,m} \nabla^2 
 + V_{B} - \mu_{B} 
 \\
 &&
 + U_{BB} \, \betrag{\Psi_{B}}^2  + U_{AB} \betrag{\Psi_{A}}^2 \Bigg] \Psi_{B} 
 +\,\lambda \, \Psi_{A} \, ,
 \nonumber
\end{eqnarray}
where $V_{A/B}$ denotes the two external potentials, and
$\mu_{A/B}$ the two chemical potentials~\cite{Garcia-Ripoll:dec2003,Jenkins-Kennedy:nov2003}.  We note that the parameter $\lambda$ can be either positive or negative without restriction, 
while the interaction parameters $U_{AA}$, $U_{AB}$, and $U_{BB}$ are typically though not always positive.
Adopting the Madelung representation, the two condensate wave-functions $\Psi_X$ can be described in terms of their densities $\{\rho_{A},\rho_{B}\}$ and phases
$\{\theta_{A},\theta_{B}\}$:
\begin{equation} \label{macrostate}
\Psi_{X}= \sqrt{\rho_{X} }\; e^{i\,\theta_{X}} \quad\hbox{for}\quad X=A,B \, .
\end{equation}
These four variables in general depend on both time and space.

\section{Wave equation}

We study zero sound in the overlap region of the
two-component system, produced by exciting density perturbations which
are small compared to the density of each condensate cloud.  In the
first experiment studying localized excitations in a one-component
BEC \cite{Ketterle:28july1997}, the optical dipole force
of a focused laser beam was used to modify the trapping potential, generating
a small density modulation. Using
phase-contrast imaging it was shown that the resulting perturbation
corresponds to a sound wave. The observed speed of sound is
\begin{equation} \label{c}
c(r)=\sqrt{\frac{4 \pi \, \hbar^2 \, a \, \rho_0(r) }{m^2}}=\sqrt{\frac{U \, \rho_0(r)}{m} }\, ,
\end{equation} 
where $\rho_0(r)$ is the density of the ground state, $a$ is the
scattering length, $m$ is the atomic mass and $U$ is the
self-interaction constant.  The mathematical equations describing
these perturbation lead to the well-known hydrodynamic equations,
which are the basis for the most fruitful of the analogies between
condensed matter physics and general relativity~\cite{Unruh,Garay,Ergo,book}.

An extension of this method can be used to obtain the kinematic equations for
small perturbations propagating in a two-component system.  Given that
the density modulation is small, the perturbations in the densities
and phases can be expanded around their macroscopic
states (\ref{macrostate}) using perturbation theory:
\begin{equation}
\Psi_{X}= \sqrt{\rho_{X0}+ \varepsilon \, \rho_{X1} }\, e^{i(\theta_{X0}+  \varepsilon \, \theta_{X1} )} 
\quad\hbox{for}\quad X=A,B \, .
\end{equation}
These states still satisfy the coupled Gross--Pitaevskii equation. When developing a perturbation analysis for the fluctuations we find that unless we set the background phases equal to each other ($\theta_{A0}=\theta_{B0}$) the calculation becomes quite intractable. Specifically, one encounters damping and damping terms dependent on $\Delta_{AB} = \theta_{A0}-\theta_{B0}$. In the appendix
(\ref{arbitraryphases}) we briefly present the result obtained for arbitrary --- even time-dependent --- background phases. 
While these terms and their implications are of interest in their own right, 
in the following focus will be set on
$\Delta_{AB}=0$ automatically implying, in particular, that the background velocities of the condensates, 
\begin{equation}
\vec{v}_{A0}=({\hbar}/{m}) \nabla \theta_{A0} 
\quad \hbox{and} \quad
\vec{v}_{B0}=({\hbar}/{m}) \nabla \theta_{B0},
\end{equation}
are equal:
 \begin{equation}
\vec  v_0 = \vec v_{A0}=\vec v_{B0}.
 \end{equation}

After a straightforward calculation, the terms of first order in
$\varepsilon$ include two coupled
equations for the perturbation of the phases
\begin{equation} \label{soundwave1}
\begin{split}
\dot{\theta}_{A1}=
&-\vec{v}_{0} \cdot \nabla \theta_{A1}  
- \frac{\tilde{U}_{AA}}{\hbar} \, \rho_{A1} 
-\frac{\tilde{U}_{AB}}{\hbar} \, \rho_{B1}, 
\\
\dot{\theta}_{B1}=
&-\vec{v}_{0} \cdot \nabla \theta_{B1}  
- \frac{\tilde{U}_{BB}}{\hbar} \, \rho_{B1} 
-  \frac{\tilde{U}_{AB}}{\hbar} \, \rho_{A1}.  \\
\end{split} 
\end{equation}
Here
\begin{eqnarray}
\tilde{U}_{AA} &=&   U_{AA} -\frac{\lambda}{2}  \frac{\sqrt{\rho_{B0}}}{(\rho_{A0})^{3/2}} ,
\nonumber
\\
\tilde{U}_{BB} &=&   U_{BB} -\frac{\lambda}{2}  \frac{\sqrt{\rho_{A0}}}{(\rho_{B0})^{3/2}} ,
\\
\tilde{U}_{AB} &=& U_{AB}+ \frac{\lambda}{2}  \; \frac{1}{\sqrt{\rho_{A0} \, \rho_{B0} }},
\nonumber
\end{eqnarray}
are modified interaction potentials for the two coupled condensates.
In addition to these two phase equations, there are two coupled
equations for the density perturbations
\begin{equation} \label{soundwave2}
\begin{split}
\dot{\rho}_{A1}=&
- \nabla\left(\frac{\hbar}{m_{A}} \,\rho_{A0} \nabla \theta_{A1} + \rho_{A1}   \;\vec{v}_{0}\right) 
\\
&
+ \frac{2\lambda}{\hbar}\sqrt{\rho_{A0} \, \rho_{B0}} \; (\theta_{B1}-\theta_{A1}), \\
\dot{\rho}_{B1}=&
- \nabla\left(\frac{\hbar}{m_{B}} \, \rho_{B0} \nabla \theta_{B1} + \rho_{B1}   \;\vec{v}_{0}\right) 
\\
&
+ \frac{2\lambda}{\hbar}\sqrt{\rho_{A0}\,\rho_{B0}} \; (\theta_{A1}-\theta_{B1}) .\\
\end{split} 
\end{equation}
To adopt a more compact representation of the physics, it is useful to define several matrices and vectors. First, define the coupling matrix
\begin{equation}
\Xi=\frac{1}{\hbar}
\left[
\begin{array}{rr}
\tilde{U}_{AA} &  \tilde{U}_{AB}  \\
\tilde{U}_{AB} & \tilde{U}_{BB}  \\
\end{array}
\right] \, .
\end{equation} 
A second coupling matrix, defined as
\begin{equation}
\Lambda=\frac{2\lambda\; \sqrt{\rho_{A0}\,\rho_{B0}} }{\hbar}
\left[
\begin{array}{rr}
+1 & -1 \\ -1 & +1 \\
\end{array}
\right] \, ,
\end{equation}
vanishes completely if the laser coupling $\lambda$ is switched off.  Last
but not least, it is also useful to introduce the mass-density matrix $D$
\begin{equation}
D={\hbar\over m}
\left[
\begin{array}{rr}
{\rho_{A0}}  & 0 \\ 0 &  {\rho_{B0}}   \\
\end{array}
\right].
\end{equation}
Now define the two-component column vectors
\begin{equation}
\bar \theta = \left[ \begin{array}{c} \theta_{A1}\\ \theta_{B1} \end{array} \right]
\qquad \hbox{and} \qquad
\bar \rho = \left[ \begin{array}{c} \rho_{A1}\\ \rho_{B1} \end{array} \right]
\end{equation}

Collecting terms into a $2\times2$ matrix equation, the equations for
the phases (\ref{soundwave1}) and densities (\ref{soundwave2}) become a compact pair of first-order PDEs
\begin{equation} \label{thetavecdot}
\dot{\bar{\theta}}=  -\,\Xi \; \bar{\rho} - \vec v_0  \cdot \nabla \bar{\theta},
\end{equation} 
\begin{equation} \label{rhovecdot}
\dot{\bar{\rho}}= \, - \nabla \cdot \left( D \; \nabla \bar{\theta} +  \bar{\rho} \; \vec{v_0} \right) 
- \Lambda \;\bar{\theta} \, .
\end{equation} 

Equation (\ref{thetavecdot})  can be used to
eliminate $\bar{\rho}$ and $\dot{\bar{\rho}}$ in equation (\ref{rhovecdot}), leaving us with a single  second-order $2\times2$ matrix  equation for
the perturbed phases:
\begin{eqnarray} \label{phaseequation}
 &\partial_{t} \left(\Xi^{-1} \; \dot{\bar{\theta}} \, \right) =
 - \partial_{t} \left(\Xi^{-1} \; \vec v_0 \cdot \nabla \bar{\theta} \, \right) 
 - \nabla    \cdot   \left(\vec v_0 \; \Xi^{-1} \; \dot{\bar{\theta}} \, \right)  
 \nonumber
 \\
&\qquad 
 + \nabla \cdot \left[ \left(D - \vec v_0 \; \Xi^{-1} \; \vec v_0 \,\cdot\, \right) \nabla \bar{\theta} \, \right] 
 + \Lambda \; \bar{\theta}.  
\end{eqnarray}
This equation tells us how a localized collective excitation in a $\lambda$-coupled
two-component BEC 
develops in time. It is a special case of the ``normal mode'' formalism developed in~\cite{normal,normal2,birefringence}.  

If we adopt a $n$-dimensional ``spacetime'' notation by writing $x^a=(t,x^i)$, with $i\in\{1,2,\cdots,n-1\}$ and $a\in\{0,1,2,\cdots,n-1\}$, then this equation can be very compactly rewritten as~\cite{normal,normal2}:
\begin{equation}
\label{E:fab}
\partial_a \left( f^{ab} \; \partial_b  \bar\theta \right) + \Lambda \;\bar \theta = 0.
\end{equation}
Here $f^{ab}$ is a $n$-dimensional rank 2 tensor density, each of whose components is a $2\times2$ matrix. Specifically
\begin{equation}
f^{00}=-\Xi^{-1}; \quad f^{0i} = - \Xi^{-1} \; v_0^i = f^{i0};
\end{equation}
and
\begin{equation}
f^{ij} = D\;\delta_{ij} -  \Xi^{-1} \; v_0^i \; v_0^j.
\end{equation}
So far there are no significant restrictions on the background densities $(\rho_{A0},\rho_{B0})$,  the joint background flow velocity $\vec v_0$, the 
interaction constants
$(U_{AA},\, U_{BB},\, U_{AB})$, and coupling constant, $\lambda$.  
If we do not decouple the two modes, then the most we can say is that the coupled system exhibits ``refringence'' in the sense of~\cite{normal2, birefringence}.

\section{Mode decoupling}

The first step in analyzing equation (\ref{phaseequation}), or the equivalent (\ref{E:fab}), is to ask
whether it is possible to tune the system so as to completely decouple it into two independent
phonon modes. Only if the two modes decouple is it possible to assign individual masses and spacetime geometries to the decoupled modes~\cite{normal, normal2, birefringence}. In the absence of decoupling, one simply has a complicated two-component system with no clear spacetime interpretation.
In performing the analysis we have found that decoupling is not possible without
introducing several constraints on the background quantities.

The decoupling analysis can be performed in several different ways, all ultimately leading to qualitatively similar physics, with minor technical differences. The major decision to be made is whether one imposes decoupling at the level of physical acoustics (at the level of the wave equation) or at the level of geometrical acoustics (at the level of dispersion relations).  The fact that physical acoustics leads to propagation phenomena more subtle than those detectable in the geometric acoustics limit is well known~\cite{Ergo,  review2}. A particularly illustrative example is provided by acoustic propagation in a fluid with non-zero vorticity~\cite{vorticity}, where the geometric acoustics approximation leads directly to a conformal class of effective spacetime metrics, while the physical acoustic approximation leads to a complicated system of PDEs.  If the only thing you can detect experimentally is the dispersion relation, then one should adopt geometrical acoustics and not demand to decouple the wave equation itself. On the other hand, if one has experimental probes that couple directly to the wave itself, then the decoupling should be performed at the level of  physical acoustics. We shall do both, and compare results later.

\section{Physical acoustics}

At the level of physical acoustics one treats the wave equation (\ref{E:fab}) as primary, then decoupling requires that all the (symmetric) matrices $f^{ab}$, and the (symmetric) matrix $\Lambda$, be simultaneously diagonalizable by position-independent orthogonal matrices $O$. That is, decoupling requires
\begin{equation}
f^{ab} = O^T \; f^{ab}_\mathrm{diag} \; O;
\quad
\Lambda = O^T \; \Lambda_\mathrm{diag} \; O; 
\quad
\tilde\theta = O \; \bar\theta;
\end{equation}
since then equation (\ref{E:fab}) becomes
\begin{equation}
\label{E:fab3}
\partial_a \left( f^{ab}_\mathrm{diag} \; \partial_b  \tilde\theta \right) 
+ \Lambda_\mathrm{diag} \;\tilde \theta= 0.
\end{equation}
Now since 
\begin{equation}
\Lambda \propto
\left[
\begin{array}{rr}
+1 & -1 \\ -1 & +1 \\
\end{array}
\right] \, ,
\end{equation}
the matrix that diagonalizes it is clearly position-independent, and the condition for simultaneous diagonalization reduces to
\begin{equation}
[f^{ab},f^{cd}]=0; \qquad [f^{ab},\Lambda]=0;
\end{equation}
where the commutators are to be interpreted in the sense of $2\times2$ matrix multiplication.  That is, the matrices  $\Xi$, $D$, and $\Lambda$ must all be simultaneously diagonalizable.

We could now proceed by direct calculation of the three commutators
\begin{equation}
[\Xi,D]; \qquad [D,\Lambda]; \qquad [\Lambda,\Xi].
\end{equation}
A perhaps more direct analysis can be performed directly in terms of equation  (\ref{phaseequation}).
Focusing on the last term in equation (\ref{phaseequation}), the
eigenvectors for non-zero coupling $\lambda \neq 0$ are given by
$\{\,[1,1]^T,\; [-1,1]^T\,\}$. The corresponding eigenvalues are $\{0,{4 \,
  \lambda \sqrt{\rho_{A0}\rho_{B0}}}/{\hbar}\}$. These eigenvectors (though not the eigenvalues) are
fixed, position-independent, and indeed independent of any of the other physical variables.  As a
result the only way to decouple equation (\ref{phaseequation}) into
two independent phonon modes, in a position-independent manner, is to demand:
\begin{equation}  \label{eigenstates}
\bar{\theta}= 
\tilde{\theta}_{1} \left[\begin{array}{r} 1 \\ 1 \end{array} \right]
+
\tilde{\theta}_{2} \left[\begin{array}{r} -1 \\ 1 \end{array} \right] \, .
\end{equation} 
We now analyze equation (\ref{phaseequation}) term by term with
respect to this particular decomposition.

Firstly, the term on the LHS, and the first two terms on the RHS in equation
(\ref{phaseequation}), have the same eigenvectors as equation
(\ref{eigenstates}) \emph{if and only if} $\tilde{U}_{AA}=\tilde{U}_{BB}$.
The eigenvalues of $\Xi^{-1}$ corresponding to the eigenvectors $\{\,[1,1]^T,\; [-1,1]^T\,\}$ are then
\begin{equation}
\left\{ 
{\hbar \over (\tilde{U}_{AA}+\tilde{U}_{AB})},
{\hbar \over
(\tilde{U}_{AA}-\tilde{U}_{AB}) }
\right\}. 
\end{equation}
This places another constraint
on the interaction variables: $\tilde{U}_{AA} \neq \pm \tilde{U}_{AB}$. [\,Indeed, note what happens if this condition fails and $\Xi$ is singular.  Then equation (\ref{thetavecdot}) cannot be solved for the column vector $\rho$ and we cannot even set up the wave equations (\ref{phaseequation}) or (\ref{E:fab}).\,]
All in all we must have
\begin{equation}
\Xi = {1\over\hbar} \left[
\begin{array}{rr}
\tilde U_{AA} & \tilde U_{AB}\\
\tilde U_{AB} & \tilde U_{AA}
\end{array}
\right]; 
\qquad
\det(\Xi)\neq 0.
\end{equation}

Secondly, we are now left with the penultimate term in equation
(\ref{phaseequation}).  Because the eigenvectors of $\Xi^{-1}$ are 
already known, there is only the
mass-density matrix $D$ to consider. 
The final constraint to decouple
the equation for the two phase perturbations  is now easily seen to be $\rho_{A0}=\rho_{B0}$. We shall simply denote this common density by $\rho_0$. That is
\begin{equation}
D=d\; \mathbf{I} = {\hbar \rho_0\over m} \; \mathbf{I}.
\end{equation}

In view of this last constraint the first coupling matrix simplifies considerably (and this was certainly not obvious when we started the analysis). We now have
\begin{eqnarray}
\tilde{U}_{AA} &=&   U_{AA} -\frac{\lambda}{2\rho_0},
\nonumber
\\
\tilde{U}_{BB} &=&   U_{BB} -\frac{\lambda}{2\rho_0},
\\
\tilde{U}_{AB} &=& U_{AB}+ \frac{\lambda}{2\rho_0},
\nonumber
\end{eqnarray}
and the first mode-decoupling constraint reduces to $U_{AA}=U_{BB}$. In order for the eigenvalues of $\Xi^{-1}$ to be well defined we need both $U_{AA}+ U_{AB}\neq0$ and $\lambda\neq\rho_0(U_{AA}-U_{AB})$, which is a mild easily satisfiable constraint.

\section{Bi-metricity}

Applying all this to equation (\ref{phaseequation}) one now sees how mode decoupling is equivalent to bi-metricity: One
obtains two [dimension-independent]
decoupled equations for the phonon modes described by the eigenstates of equation
(\ref{eigenstates}):
\begin{equation}
\partial_a (f^{ab}_I \;\partial_b \tilde\theta_I) = -
{4\lambda\rho_0\over \hbar} \; \delta_{2I}\; \tilde\theta_I,
\quad\hbox{for}\quad
I=1,2,
\end{equation}
where $\delta_{2 \, I}$ is the usual Kronecker delta.
Here
\begin{equation}
f^{ab}_I={d\over c^2_I}
\left[
\begin{array}{c|c}
-1         & -v_0^j \\
\hline
-v_0^j & c_I^2\;  \delta^{ij} - v_0^i \; v_0^j
\end{array}\right] \, ,
\end{equation}
where the possibly distinct propagation speeds $c_I$ are defined in terms of the eigenvalues $\Xi_I$ of the matrix $\Xi$ by
\begin{equation} 
\label{cis}
c_{I}^2=   \Xi_I \; d =  \frac{d \, (\tilde{U}_{AA} + (-1)^{I} \tilde{U}_{AB})}{\hbar} \, ,
\end{equation}
that is
\begin{equation} 
\label{cis2}
c_{I}^2=  \frac{\rho_0 \, (\tilde{U}_{AA} + (-1)^{I} \tilde{U}_{AB})}{m} \, .
\end{equation}
It is important to note that decoupling by itself does not force the two propagation speeds to be equal --- decoupling in this context generically produces a two-metric model [bi-metricity], and the demand that every phonon couple to a single effective metric [mono-metricity] is an additional independent condition.  (In general the phonons arising from a system of $N$ interacting BECs would, if they decoupled in the above manner, lead to an $N$-metric model.) 

Introducing the (dimension-independent) natural oscillation frequency
\begin{equation}
\omega_0^2 
= -
{4\lambda \, \rho_0 \; c_2^2 \over \hbar\; d} 
= -
{4\lambda \, m \; c_2^2\over \hbar^2},
\end{equation}
we can write
\begin{equation}
\partial_a (f^{ab}_I \;\partial_b \tilde\theta_I) = {d\over c_2^2} \; \omega_0^2 \; \delta_{2I}\; \tilde\theta_I,
\quad\hbox{for}\quad
I=1,2,
\end{equation}
where $\omega_0$ now has the physical interpretation that it is the frequency at which a position-independent (zero-momentum) configuration oscillates.

When converting the contravariant tensor densities $f^{ab}_I$ into covariant spacetime metrics $g_{ab}$ one encounters a number of dimension-dependent factors~\cite{Ergo,vortex}, such that naive application of the formalism leads to peculiar dimensionalities for physical quantities. The best way we have found for keeping track of these factors is to introduce space and time independent \emph{reference constants} $c_*$ and $d_*$ and write
\begin{equation}
\tilde f^{ab} = {c_*\over d_*} \; f^{ab} = \left({c_* \;d\over c_I\; d_*}\right) \; {1\over c_I}
\left[
\begin{array}{c|c}
-1         & -v_0^j \\
\hline
-v_0^j & c_I^2\;  \delta^{ij} - v_0^i \; v_0^j
\end{array}\right] \, ,
\end{equation}
so that
\begin{equation}
\partial_\mu (\tilde f^{\mu\nu}_I \;\partial_\nu \tilde\theta_I) = {c_* \; d\over c_2\; d_*} \; {\omega_0^2 \; \delta_{2I}\over c_2} \; \tilde\theta_I,
\quad\hbox{for}\quad
I=1,2.
\end{equation}
Here $c_*$ and $d_*$ are any convenient but fixed reference values. They might be (for instance) the spatial average values of $c_2$ and $d$  over the entire condensate, or they might be chosen in terms of the speed of light and other fundamental constants. They are introduced for convenience, and do not ultimately affect the physics we are discussing.

Now introducing a pair of effective ``spacetime metrics'' by the identifications
\begin{equation}
\sqrt{-g_I} \; g_I^{ab} = \tilde f_I^{ab}
\quad\hbox{and}\quad  
g_I = {1\over\det[g_I^{ab}]}, 
\end{equation}
we can recast these wave equations equations as a pair of
curved-space Klein-Gordon [massive d'Alembertian] equations
\begin{equation} \label{Klein-Gordon equation}
\frac{1}{\sqrt{-g_I}}
\partial_{a} \left( \sqrt{-g_I} \, g^{ab}_{I} \partial_{b} \, \tilde{\theta}_{I} \right) 
= {\mathbf{m_{\mathrm{phonon}}^2} \; c_*^2\over\hbar^2} \,  
\delta_{2\, I} \,  \tilde{\theta}_{I},
\end{equation} 
where as we shall soon see all quantities carry their standard dimensionality. 

After a brief algebraic calculation we find
\begin{equation}
\sqrt{-g_I} = c_I \left( {c_*\; d\over c_I \; d_*} \right)^{n/(n-2)}
\end{equation}
and so
\begin{equation}
g^{ab}_I=  \left({c_*\;d\over c_I\;d_*}\right)^{-n/(n-2)} 
\left\{  {1\over c^2_I} 
\left[
\begin{array}{c|c}
-1         & -v_0^j \\
\hline
-v_0^j & c_I^2 \delta^{ij} - v_0^i \; v_0^j
\end{array}\right] \right\}.
\end{equation}
These quantities depend on the space-time dimension $n$~\cite{Ergo,vortex} in
such a manner that
\begin{equation}
g_{ab}^I=  \left({c_*\;d\over c_I\;d_*}\right)^{n/(n-2)} \; 
\left[
\begin{array}{c|c}
-(c_I^2 - v_0^2)         & -v_0^j \\
\hline
-v_0^j & \delta^{ij}
\end{array}\right]\, .
\end{equation}
Finally the mass-term is
\begin{equation}
\mathbf{m_{\mathrm{phonon}}^2}=  {\hbar^2\;\omega_0^2\over c_*^2\;c_2^2} 
\;
\left({c_*\;d\over c_2\;d_*}\right)^{-2/(n-2)}.
\end{equation}
We note that this ``mass'' term can in principle depend on position, so some authors might prefer to refer to it as a ``potential'' term. There is no universal agreement on  terminology regarding this point, but the bulk of the community would be happy to refer to this as a ``position dependent mass''.  Furthermore, if one focuses on the core of the BEC cloud, where all gradients are by definition zero (or small), then this mass term is guaranteed to be approximately constant.

Now the two metrics $g_{ab}^I$, the inverse metrics $g^{ab}_I$, and the phonon mass $\mathbf{m_{\mathrm{phonon}}^2}$, all depend on the normalization constants $c_*$ and $d_*$. This is as it should be, since $c_*$ and $d_*$ were introduced to give canonical dimensions to these quantities. In particular, since $c_*$ and $d_*$ contribute to the overall conformal factor in the analogue spacetime metric, they set the overall scale with respect to which analogue ``masses'' are to be measured. However, correctly formulated physical questions will depend only on parameters such as $\omega_0$, $c_I$, and $v_0$ which are independent of these normalization constants. 
For instance, in the eikonal limit the dispersion relation for these two decoupled phonon modes is simply
\begin{equation}
\label{fresnel-2-decoupled}
\left(\omega - \vec{v}_0 \cdot \vec{k} \right)^2 - c_I^2 \; k^2 = \omega_0^2\; \delta_{2I}\,.
\end{equation}
(A similar calculation, but restricted to a one-condensate system,
where all variables are likewise allowed to be time and space dependent, but no
mass term is present, has been presented in \cite{silke2}.)

For the phase vector $\tilde{\theta}_{1}$, (corresponding to perturbations in the two
condensates $A$ and $B$ oscillating ``in phase''), the mass term is
always zero. However, for a laser-coupled system ($\lambda\neq0$) the
mass-term in the equation for $\tilde{\theta}_{2}$, (corresponding to
perturbations in the two condensates $A$ and $B$ oscillating in
``anti-phase''), does not vanish.

\section{Mono-metricity}
 Comparing the definition for the speed of sound (\ref{c}) in a one
component system, with the two speeds $c_I$ introduced here, we see that the
$c_{I}$ (\ref{cis}) are simply the $\lambda$-modified speeds of sound for each decoupled phonon
mode. 
(For instance, consider an idealized situation in which the two condensates decouple completely, $U_{AB}=0$ and $\lambda=0$, the
two $c_I$'s become the independent phonon speeds in each condensate
cloud.)
This fact leaves us with the possibility of constructing two
different types of analogue model. So far we have been dealing with a
two-metric structure, which is interesting in itself~\cite{normal2,vsl}. For instance, in
the absence of laser-coupling, $\lambda=0$,  the presence of two
different speed of sounds can be used for tuning
effects~\cite{Uwe:new}.

If we wish to more accurately simulate the curved spacetime of our own
universe, another constraint should be placed on the system, to make
the two speeds of sound equal $c=c_{1}=c_{2}$.  This yields a single
sound-cone structure, to match the observed fact that our universe
exhibits a single light-cone structure.  This condition is fulfilled
if we set 
\begin{equation}
\tilde{U}_{AB}=0; \quad\implies\quad \lambda = - 2 \;U_{AB}\;\rho_0.
\end{equation} 
In this case
\begin{equation}
c^2 = \frac{\rho_0\;\tilde U_{AA}}{m} = \frac{\rho_0\;(U_{AA} + U_{AB})}{m},
\end{equation}
while we have the dimension-independent result that the natural oscillation frequency becomes
\begin{equation}
\omega_0^2 = 8\;\frac{\rho_0^2\;U_{AB}(U_{AA} + U_{AB})}{\hbar^2}.
\end{equation}
In counterpoint
\begin{equation}
\mathbf{m_{\mathrm{phonon}}^2} = {8\;U_{AB} \;\rho_0 \; m\over c_*^2} 
\left[ { (U_{AA}+U_{AB}) \;d_*^2\; m
\over
\hbar^2 \;c_*^2\;\rho_0} \right]^{1/(n-2)}.
\end{equation}

We note that in typical situations $U_{AA} \approx U_{BB}$ and $U_{AB}$ are both positive, corresponding to repulsive atomic interactions~\cite{Jenkins-Kennedy:nov2003}.
This implies that $\lambda$ is then negative, corresponding to a negative trapping potential, but a positive $\omega_0^2$ and a positive $\mathbf{m}_\mathrm{phonon}^2$.  

It is also possible to choose systems such that $U_{AB}$ is negative, as long as $c^2$ is still kept positive, which in turn requires $U_{AA}+U_{AB}$ to be positive, and places a restriction on the relative sizes of the self-interactions and cross-interactions.
This implies that $\lambda$ is positive.  
Furthermore 
the natural oscillation frequency $\omega_0^2$ and the phonon mass $\mathbf{m_{\mathrm{phonon}}}^2$ will then be real and negative.  That is, attractive atomic interactions, which signal an instability in the condensate, would in our analysis correspond to a tachyonic phonon. 

If in contrast we permit $c^2$ to go negative (that is we permit  $U_{AA}+U_{AB}$ to be negative) then we have a ``Euclidean signature regime'' where phonons do not propagate --- this corresponds to the gross instability of the condensate which manifests itself as the ``Bose-Nova'' phenomenon~\cite{Bei-Lok-Hu}.

In this mono-metric situation, while the in-phase perturbations will propagate
exactly at the speed of sound, 
\begin{equation}
\vec{v}_{s} =\vec{v}_0 + \hat k \;c\, ,
\end{equation}
the anti-phase perturbations will move
with a lower group velocity given by:
\begin{equation}
\vec{v}_{g}= \frac{\partial \omega}{\partial \vec{k}}
=\vec{v}_0 + \hat k \;\frac{c^2}{\sqrt{\omega_0^2\; + c^2\; k^2}}
\, .
\end{equation}
Here $k$ is the usual wave number. This explicitly
demonstrates that the group velocity of the anti-phase eigenstate
depends on the laser-induced coupling between the condensates.

\section{Special case: Constant $\Xi$}
\def\tr{\mathrm{tr}}

There is a special case that is worth considering separately. Suppose that the matrix $\Xi$ is a time and space-independent constant. Then by defining
\begin{equation}
\bar\theta_\mathrm{new} = \Xi^{-1/2}\; \bar\theta,
\end{equation}
multiplying equation (\ref{thetavecdot}) by $\Xi^{+1/2}$, and appropriately commuting $\Xi$ through the space and time derivatives, one can re-write equation (\ref{thetavecdot}) as
\begin{eqnarray} \label{phaseequation2}
 &\partial_{t}^2\bar{\theta}_\mathrm{new} =
 - \partial_{t} \left(\mathbf{I} \; \vec v_0 \cdot \nabla \bar{\theta}_\mathrm{new} \, \right) 
 - \nabla    \cdot   \left(\vec v_0 \; \mathbf{I} \; \dot{\bar{\theta}}_\mathrm{new} \, \right)  
 \nonumber
 \\
&\qquad 
 + \nabla \cdot \left[ \left(C^2 - \vec v_0 \; \mathbf{I} \; \vec v_0 \,\cdot\, \right) \nabla \bar{\theta}_\mathrm{new} \, \right] 
 - M^2 \; \bar{\theta}_\mathrm{new},
\end{eqnarray}
where
\begin{equation}
C^2 = \Xi^{1/2}\; D \;\Xi^{1/2}; \qquad M^2 =   -\Xi^{1/2} \;\Lambda\; \Xi^{1/2}.
\end{equation}
Now the existence of the matrix square root $\Xi^{1/2}$ follows since $\Xi$ itself is real and symmetric. Indeed by the Hamilton--Cayley theorem for $2 \times 2$ matrices we know
\begin{equation}
\Xi^{1/2} = a \, \mathbf{I} + b \, \Xi ,
\end{equation}
where $a$ and $b$ are functions of the eigenvalues of $\Xi$. A little matrix algebra yields (for $2\times2$ matrices only) the explicit formula
\begin{equation}
\Xi^{1/2} = \frac{\sqrt{\det \Xi} \;\; \mathbf{I} + \Xi }{\sqrt{2  \sqrt{\det \Xi} + \tr( \Xi)}} .
\end{equation}
It is now clear that both $C^2$ and $M^2$ are symmetric matrices, so in this constant-$\Xi$ special case mode decoupling of the wave equation (\ref{phaseequation2}), that is bi-metricity, requires the simple constraint
\begin{equation}
[C^2, \; M^2] = 0,
\end{equation}
which is equivalent to the matrix equation
\begin{equation}
D \; \Xi \; \Lambda = \Lambda \; \Xi \; D.
\end{equation}
In terms of the physical parameters we obtain the constraint
\begin{equation} 
\label{decoup-const-Xi}
\rho_{A0} \, \tilde{U}_{AA} - \rho_{B0} \, \tilde{U}_{BB} = (\rho_{A0} - \rho_{B0}) \, \tilde{U}_{AB}.
\end{equation}
Note that to get to this stage we had to assume $\Xi$ was constant (so $\tilde U_{AA}$, $\tilde U_{AB}$, and $\tilde U_{BB}$ are constant), and then from the above we see that decoupling requires that $\rho_{A0}/\rho_{B0}$ must be a position and time-independent constant. We do not however need $\rho_{A0}=\rho_{B0}$; by \emph{restricting} the position and time dependence of $\Xi$ 
we have permitted other parts of the wave-equation to possess an algebraically more \emph{general} solution to the decoupling constraint (the bi-metricity constraint). Subject only to the condition (\ref{decoup-const-Xi}) the speeds of sound and masses are simply extracted as eigenvalues of the matrices $C^2$ and $M^2$ respectively.

Imposing the bi-metricity condition (\ref{decoup-const-Xi}) the two eigenvalues of $C^2$, the two speeds of sound, are
\begin{equation}
\label{e:csq-Xi}
c_{\pm}^2 = \frac{1}{2 m} \left\{ \rho_{A0} \tilde{U}_{AA} + \rho_{B0} \tilde{U}_{BB} \pm (\rho_{A0} + \rho_{B0} ) \, \tilde{U}_{AB} \right\}.
\end{equation}
Additionally imposing the mono-metricity condition requires the two eigenvalues
of $C^2$ to be the same. That is, mono-mericity enforces $\tilde{U}_{AB} = 0$.
Together with the decoupling
condition equation (\ref{decoup-const-Xi}), we now get:
\begin{equation}
\label{e:mono-Xi}
\rho_{A0} \; \tilde{U}_{AA} = \rho_{B0} \; \tilde{U}_{BB}.
\end{equation}
That is
\begin{equation}
c^2 = {\rho_{A0} \; \tilde U_{AA}\over m} = {\rho_{B0} \; \tilde U_{BB}\over m}.
\end{equation}
and
\begin{equation}
\begin{split}
\omega_0^2 &= \tr[M^2] \\
&= {4 \, \rho_{A0} \, \rho_{B0} \, U_{AB} 
\; \left[ U_{AA}+ U_{BB} + U_{AB} \left( \frac{\rho_{A0}}{\rho_{B0}} + \frac{\rho_{B0}}{\rho_{A0}}  \right) \right] \over \hbar^2} \\
\end{split}
\end{equation}
The eigenvector corresponding to the massless phonon mode is  $ \bar\theta_\mathrm{new} \propto [\sqrt{\tilde U_{BB}}, \; \sqrt{\tilde U_{AA}} ]^T$, while that for the massive phonon mode is $ \bar\theta_\mathrm{new} \propto [-\sqrt{\tilde U_{AA}}, \; \sqrt{\tilde U_{BB}} ]^T$. In terms of the original variables $\bar\theta$ this corresponds to decoupled eigenmodes  $\bar\theta \propto [1, \; 1 ]^T$ for the massless mode, while that for the massive phonon mode $ \bar\theta \propto [-\tilde U_{AA}, \; \tilde U_{BB} ]^T$. 
%

\section{Special case: Constant $D$}

Similarly, consider the case where the matrix $D$ is independent of position and time --- this corresponds to a situation where the background densities $\rho_{A0}$ and $\rho_{B0}$ are constant, though they do not need to be equal. In this situation we can define
\begin{equation}
\bar\theta_\mathrm{new} = D^{+1/2}\; \bar\theta.
\end{equation}
Then multiplying equation (\ref{thetavecdot}) by $D^{-1/2}$, and appropriately commuting $D$ through the space and time derivatives, one can re-write equation (\ref{thetavecdot}) as
\begin{eqnarray} \label{phaseequation3}
 &\partial_{t}( C^{-2} \partial_t \bar{\theta}_\mathrm{new}) =
 - \partial_{t} \left(C^{-2} \; \vec v_0 \cdot \nabla \bar{\theta}_\mathrm{new} \, \right) 
 \qquad\qquad
 \nonumber\\
 & \qquad
 - \nabla    \cdot   \left(\vec v_0 \; C^{-2} \; \dot{\bar{\theta}}_\mathrm{new} \, \right)  
 \nonumber
 \\
&\qquad 
 + \nabla \cdot \left[ \left(\mathbf{I} - \vec v_0 \; C^{-2} \; \vec v_0 \,\cdot\, \right) \nabla \bar{\theta}_\mathrm{new} \, \right] 
 - \tilde M^2 \; \bar{\theta}_\mathrm{new},\quad
\end{eqnarray}
where now
\begin{equation}
C^{-2} = D^{-1/2} \; \Xi^{-1}\; D^{-1/2}; 
\qquad 
\tilde M^2 =   - D^{-1/2} \;\Lambda\; D^{-1/2}.
\end{equation}

The analysis is now similar to the case of constant-$\Xi$.  Decoupling (bi-metricity) requires
\begin{equation}
[ C^{-2}, \tilde M^2] = 0,
\end{equation}
that is
\begin{equation}
\Xi^{-1}\; D^{-1} \; \Lambda = \Lambda \; D^{-1} \; \Xi^{-1}.
\end{equation}
It is easy to rearrange this equation to yield
\begin{equation}
\Lambda\; \Xi\; D = D\; \Xi\; \Lambda,
\end{equation}
which is the \emph{same} matrix equation as encountered in the  constant-$\Xi$ case. 
Consequently, decoupling requires the \emph{same} algebraic condition, equation (\ref{decoup-const-Xi}) as in the constant-$\Xi$ case. The speeds of sound are now given by the eigenvalues of the matrix
\begin{equation}
\label{e:csq-D}
C^2 = D^{+1/2} \; \Xi\; D^{+1/2},
\end{equation}
and explicit computation again yields, after imposing the decoupling constraint (\ref{decoup-const-Xi}),  the same algebraic result (\ref{e:csq-Xi}). Imposing mono-metricity again leads to $\tilde U_{AB}=0$ and equation (\ref{e:mono-Xi}), which in this case can be read off by inspection from equation (\ref{e:csq-D}). In short, there is no new physics hiding in the case of constant $D$; we again see  that by \emph{restricting} the position and time dependence of some of the coefficients in the wave equation (\ref{phaseequation})
we have permitted other parts of the wave-equation to possess a slightly more general solution to the decoupling constraint (the bi-metricity constraint).

\section{Geometrical acoustics --- Fresnel equation}

In geometrical acoustics we adopt the eikonal approximation 
\begin{equation}
\bar\theta = \mathcal{A}\; \exp(-i\varphi)
\end{equation}
with a slowly varying amplitude $\mathcal{A}$ and a rapidly varying phase $\varphi$. We also assume that the coefficients in the differential equation (\ref{phaseequation}) are slowly varying compared to the phase. This approximation leads, before we apply the decoupling constraints,  to the Fresnel
equation~\cite{normal,normal2}
\begin{eqnarray}
\label{fresnel}
&&\det\big\{
\omega^2 \; \Xi^{-1}  -  2 \omega \; (\vec v_0\cdot\vec k) \; \Xi^{-1} 
\nonumber
\\
&&
\qquad
- [ D \; k^2 -  (\vec v_0\cdot\vec k)^2\; \Xi^{-1} ] + \Lambda ]
\big\}
=0\,.
\end{eqnarray}
In a 2-component BEC the Fresnel equation is in general a quartic dispersion relation for two interacting phonon modes. This approximation makes sense when the period and wavelength of the phonon mode is small compared to the timescale and distance scale over which the background configuration changes.

As was the case in the constant-$\Xi$ physical acoustics analysis, it is convenient to pre-multiply and post-multiply the Fresnel equation by $\Xi^{+1/2}$, thereby re-writing the Fresnel equation as
\begin{eqnarray}
\label{fresnel2}
&&\det\big\{
\omega^2 \;  \mathbf{I} -  2 \omega \; (\vec v_0\cdot\vec k) \; \mathbf{I}
\nonumber
\\
&&
\qquad
- \left[C^2 \; k^2 -  (\vec v_0\cdot\vec k)^2 \; \mathbf{I} \right] - M^2 
\big\}
=0\,,
\end{eqnarray}
where again
\begin{equation}
C^2 = \Xi^{1/2}\; D \;\Xi^{1/2}; \qquad M^2 =   -\Xi^{1/2} \;\Lambda\; \Xi^{1/2}.
\end{equation}
Thus by adopting the eikonal approximation implicit in the Fresnel equation one has reduced the number of matrices one needs to deal with to $C^2$ and $M^2$, now without needing to strictly enforce the spatial and temporal constancy of $\Xi$ (or $D$).
The Fresnel equation now becomes
\begin{equation}
\label{fresnel3}
\det\left\{ \left(\omega-\vec v_0\cdot\vec k\,\right)^2 \; \mathbf{I} - C^2 k^2 - M^2 \right\} = 0.
\end{equation}

Imposing mode decoupling, which is now equivalent to the Fresnel equation factorizing into two quadratics, forces $M^2$ and $C^2$ to be simultaneously diagonalizable. [In which case we recover equation (\ref{fresnel-2-decoupled}).] Finally, imposing mono-metricity enforces $C^2=c^2 \; \mathbf{I}$, with $M^2$ symmetric and singular but otherwise unconstrained.

Thus an analysis in terms of the Fresnel equation leads to the same conclusions as the direct physical acoustics analysis in the special case of constant $\Xi$, or the special case of constant $D$  --- these are the same algebraic constraints as were obtained when analyzing equations (\ref{phaseequation2}) and (\ref{phaseequation3}), albeit in a slightly different physical regime.

%
\section{Discussion}

In conclusion, the calculation presented in this article is of
interest to two separate communities. For the BEC community, it
provides a specific example of how to tune an interacting two-BEC
condensate in such a manner as to obtain a massive phonon. With the background configurations in phase, but without the
fine tuning, it provides an example of two interacting phonon modes
whose wave equation is governed by the second-order coupled system of PDEs  (\ref{phaseequation}) or equivalently (\ref{E:fab}), and whose dispersion relation is governed by the quartic Fresnel equation
(\ref{fresnel3}). 

It is important to note that fully decoupling the wave equation in a completely general background is algebraically more restrictive than the problem of decoupling the Fresnel equation in the geometrical acoustics limit. Indeed in the geometrical acoustics limit decoupling places algebraic constraints on the coefficients of the wave equation that are equivalent to physical acoustics in the special case where the matrix $\Xi$ or the matrix $D$ is both position and time independent. This situation is to a good approximation relevant at and near the centre of the BEC cloud. The subtleties involved in implementing decoupling into independent modes is surprisingly more complex than one might at first envisage.

If for the sake of discussion we insert specific numbers relevant to a BEC mixture based on hyperfine states in $^{87}$Rb, we find $\omega_0\approx 150$ kHz.  For a harmonic magnetic trap this should be compared with a typical trap oscillation frequency of $100$ kHz, though note that for non-harmonic traps one can in principle make the trap oscillation frequency arbitrarily small. Similarly we find $\lambda \approx  10^{29} \hbox{ J}$, corresponding to a laser temperature of $800$ nK (to be compared to a BEC condensation temperature of some $\mu$K). In short, the effects we have been considering in this article lie at  the cutting edge of present day experimental technique. In this regard it is important to realise that multi-component BECs have already been constructed in the laboratory~\cite{anu}.
The technological issues in actually implementing this approach for generating the massive Klein--Gordon equation amount to keeping the background condensates in phase while decoupling the phonon modes in a simple manner. Doing this may be experimentally challenging, but there appears to be no obstacle in principle to actually implementing the model.

From the general relativity perspective, this article settles an important matter of principle: It 
provides an example of an analogue system that can be used to mimic a
minimally coupled massive scalar field embedded in a curved spacetime. 
Quantum fields of this type are essential for any realistic application of analogue spacetime ideas to particle physics, and in particular are essential for developing condensed matter simulations of quantum gravity phenomenology.

\section*{Acknowledgments}

This research was supported by the Marsden
Fund administered by the Royal Society of New Zealand.  The authors wish to thank Crispin Gardiner, Piyush Jain, Ashton Bradley, and John Close for their thoughtful comments.

\appendix
\section{Wave equation for arbitrary initial phases}
\label{arbitraryphases}

If we start our calculations with a more complex phase relationship between the two condensates then the wave 
equation (\ref{phaseequation}) gains additional terms.

The result we obtain for two arbitrary initial --- possibly even time-dependent --- background phases is:
\begin{equation} 
\label{fancyphaseequation}
\begin{split}
&
 \partial_t \, \left( \Xi^{-1} \,\dot{\bar{\theta}} \right) =  
- \partial_t \left( \Xi^{-1} \, \vec{V} \, \nabla  \bar{\theta} \right) 
- \nabla \left( \vec{V} \, \Xi^{-1} \dot{\bar{\theta}} \right) 
\\
& 
+\nabla \, \left( \left( D - \vec{V} \, \Xi^{-1} \, \vec{V}    \right) \nabla \bar{\theta} \right) 
+ \Lambda \; \bar \theta
\\
&
+ \Big\{ - C_\mathrm{density} \,  \Xi^{-1} \,C_\mathrm{phase}   
\\
& +  \partial_t \left( \Xi^{-1} \, C_\mathrm{phase}  \right) 
+ \nabla \left( \vec{V} \Xi^{-1} C_\mathrm{phase} \right)  \Big\} \; \bar{\theta} 
\\
& 
+  \left\{ \Xi^{-1} \, C_\mathrm{phase} + C_\mathrm{density} \, \Xi^{-1}  \right\} \, \dot{\bar{\theta}} 
\\
&
+  \left\{ \vec{V} \, \Xi^{-1} \, C_\mathrm{phase} + 
C_\mathrm{density} \, \Xi^{-1}  \, \vec{V}  \right\} \nabla \bar{\theta} . 
\\
\end{split} 
\end{equation} 
In addition to those matrices defined in the previous calculations, three new matrices show up.

The matrix $\vec{V}$ simply contains the two background velocities of each condensate,
\begin{equation}
\vec{V}= 
\left[
\begin{array}{cc}
\vec{v}_{A0} & 0  \\
0 & \vec{v}_{B0}  \\
\end{array}
\right] ,
\end{equation} 
now with two possibly distinct background velocities,
\begin{equation}
\begin{split}
\vec{v}_{A0} = & \frac{\hbar}{m} \nabla \theta_{A0}, \\
\vec{v}_{B0} = & \frac{\hbar}{m} \nabla \theta_{B0} ,\\
\end{split} 
\end{equation}

Additionally, we also obtain two completely new matrices, which vanish in the case of identical initial phases
\begin{equation}
\Delta_{AB} := \theta_{A0} - \theta_{B0} = 0.
\end{equation}
These new matrices are the coupling-phase matrix,
\begin{equation}
C_\mathrm{phase}= \frac{\lambda \sin \Delta_{AB}}{\hbar} 
\left[
\begin{array}{cc}
+\sqrt{\frac{\rho_{B0}}{ \rho_{A0}}} & -\sqrt{\frac{\rho_{B0}}{\rho_{A0}}}  \\
+\sqrt{\frac{\rho_{A0}}{\rho_{B0}}} & -\sqrt{\frac{\rho_{A0}}{\rho_{B0}}} \\
\end{array}
\right],
\end{equation} 
and the coupling-density matrix
\begin{equation}
C_\mathrm{density}= \frac{\lambda \sin \Delta_{AB}}{\hbar} 
\left[
\begin{array}{cc}
-\sqrt{\frac{\rho_{B0}}{ \rho_{A0}}} & -\sqrt{\frac{\rho_{A0}}{\rho_{B0}}}  \\
+\sqrt{\frac{\rho_{B0}}{\rho_{A0}}} & +\sqrt{\frac{\rho_{A0}}{\rho_{B0}}}  \\
\end{array}
\right] .
\end{equation}  
In the case of identical phases equation (\ref{fancyphaseequation}) simplifies to the
wave equation (\ref{phaseequation}) found in the previous calculation.  Specifically the first two lines of (\ref{fancyphaseequation})  are exactly the same as  (\ref{phaseequation}), while the next two lines represent a mass-shift, and the last two lines correspond to a damping-term and a smoothing term respectively.

\vfil



\end{document}